\documentclass[11pt,a4paper,twocolumn]{article}
\usepackage{graphicx}
\usepackage{rotating}
\usepackage{amssymb}
\usepackage{mathptmx}
\usepackage{upgreek}
\usepackage[numbers]{natbib}
\usepackage{authblk}
\usepackage{xcolor}
\usepackage{subfigure}

\makeatletter

\begin{document}

\title{\textbf{Susceptibility study of TES micro-calorimeters\\for X-ray spectroscopy under FDM readout}}

\author[1]{D.~Vaccaro\thanks{d.vaccaro@sron.nl}}
\author[1]{H.~Akamatsu}
\author[1]{L.~Gottardi}
\author[2]{J.~van~der~Kuur}
\author[1]{E.~Taralli}
\author[1]{M.~de~Wit}
\author[1]{M.P.~Bruijn}
\author[1]{R.~den~Hartog}
\author[3]{M.~Kiviranta}
\author[1]{A.J.~van~der~Linden}
\author[1]{K.~Nagayoshi}
\author[1]{K.Ravensberg}
\author[1]{M.L.~Ridder}
\author[1]{S.~Visser}
\author[1]{B.D.~Jackson}
\author[1,4]{J.R.~Gao}
\author[1]{R.W.M.~Hoogeveen}
\author[1,5]{J.W.A.~den~Herder}

\affil[1]{NWO-I/SRON Netherlands Institute for Space Research, Niels Bohrweg 4, 2333CA Leiden, Netherlands}
\affil[2]{NWO-I/SRON Netherlands Institute for Space Research, Landleven 12, 9747 AD Groningen, Netherlands}
\affil[3]{VTT Technical Research Centre of Finland, Tietotie 3, 02150 Espoo, Finland}
\affil[4]{Optics Group, Department of Imaging Physics, Delft University of Technology, Delft, 2628 CJ, Netherlands}
\affil[5]{Universiteit van Amsterdam, Science Park 904, 1090GE Amsterdam, The Netherlands}

\date{}

\twocolumn[
\begin{@twocolumnfalse}
\maketitle
		
\begin{quotation}
\textbf{This paper has been accepted for publication in \textit{Journal of Low Temperature Physics}.}
\end{quotation}

\begin{abstract}

We present a characterization of the sensitivity of TES X-ray micro-calorimeters to environmental conditions under frequency-domain multiplexing (FDM) readout. In the FDM scheme, each TES in a readout chain is in series with a LC band-pass filter and AC biased with an independent carrier at MHz range. Using TES arrays, cold readout circuitry and warm electronics fabricated at SRON and SQUIDs produced at VTT Finland, we characterize the sensitivity of the detectors to bias voltage, bath temperature and magnetic field. We compare our results with the requirements for the Athena X-IFU instrument, showing the compliance of the measured sensitivities. We find in particular that FDM is intrinsically insensitive to the magnetic field because of TES design and AC readout.

\end{abstract}
\end{@twocolumnfalse}
]

\section{Introduction}\label{intro}

Transition-edge sensors\cite{tes} (TES) are the baseline detector technology in future X-ray space-borne telescopes, such as Athena X-IFU\cite{athena}, Lynx\cite{lynx} and HUBS\cite{hubs}. To meet the scientific goals, thousands of TESs will be hosted on the focal plane of these instruments, with high requirements on energy resolution, spatial resolution and count-rate capability. Given the stringent limitations of space-borne missions in terms of available cooling power at cryogenic temperatures, electrical power and mass, the readout of TESs is usually performed under a multiplexing scheme, the most common ones being Time-Division Multiplexing\cite{tdm} (TDM) and Frequency-Domain Multiplexing\cite{hiroki2021} (FDM) (a description of the TES architecture, as well as both readout designs, can be found in \textit{Gottardi, Nagayoshi 2021}\cite{tes}). Requirements on the detector spectral performance dictate a stringent energy resolution budget on the various contributors to the total instrumental energy resolution, such as detector and readout noise, sensitivity to environmental conditions and instrumental drifts. In particular, the latter factors affect the total energy resolution via undesired variations of the TES responsivity.

The responsivity, or gain, of a TES to a photon of a certain energy depends on the setpoint along the superconducting transition, which is defined by the bias voltage $V$, the bath temperature $T$ and the magnetic field $B$. A small change in these parameters can affect in some measure the TES gain, which implies that photons of identical energy $E$ could generate pulses of different height and/or shape. This effect can in principle degrade the instrumental energy resolution. For this reason, gain sensitivities are important parameters to characterize, since they contribute to defining the instrumental design. For example, the $B$-field sensitivity is related to the magnetic shielding: a lower sensitivity to magnetic fields allows for a magnetic shield of lower mass, a very important factor to consider for a space-born instrument.

SRON has been developing, in the framework of Athena X-IFU, TES micro-calorimeters for X-ray spectroscopy\cite{ken} and a frequency-domain multiplexing (FDM) readout with base-band feedback (BBFB)\cite{bbfb,hiroki2021}. In the FDM scheme, the readout of a TES array is performed by placing a tuned high-$Q$ LC band-pass filter in series with each detector and providing an ac-bias with an independent carrier in the MHz range. The signals of all the detectors in the readout chain are summed at the input coil of a Superconducting QUantum Interference Device (SQUID), which provides a first amplification at cryogenic temperature. Further amplification and conversion into digital is performed at room temperature by a control electronics board, which is also responsible for the bias carrier generation and demodulation of the output comb. In the BBFB scheme, the demodulated TES signals are again remodulated using the same carrier frequencies, compensated with a phase delay and fed back at the SQUID feedback coil, to null the current at the SQUID input: this allows a more efficient use of the SQUID dynamic range and effectively increases the number of pixels readable in a single readout chain.

The FDM scheme is fundamentally different from TDM, where TESs are dc-biased. In this contribution, we present a characterization of the gain sensitivities of a TES array using a cryogenic FDM setup.

\section{Experimental setup and measurement method}\label{method}

For our experiments we use a $8\times 8$ uniform TES array, with 31 devices connected to the readout circuit. Each TES is a $80\times 13\ \upmu$m$^{2}$ Ti/Au bilayer, coupled to a $240\times 240\ \upmu$m$^{2}$, 2.3~$\upmu$m thick Au absorber (thermal capacitance $C \simeq 0.85$~pJ/K at 90~mK) via two central pillars and with four additional corner stems providing mechanical support. These devices have critical temperature $T_{C} \simeq 84$~mK, normal resistance $R_{N} \simeq 155$~m$\upOmega$ and thermal conductance $G\sim~65$~pW/K at $T_{C}$.

For the FDM readout, the TESs are coupled to custom superconducting LC filters \cite{marcel} and transformers for impedance matching. A stiff voltage bias is provided through an effective shunt resistance of $\sim 1$~m$\upOmega$. The TES summed signals are pre-amplified at cryogenic temperature via two SQUIDs (Front-End + Amplifier). Such "cold" components are hosted on custom oxygen-free high-conductivity (OFHC) copper and enclosed in a niobium shield. Superconducting Helmholtz coils are employed to control the magnetic field applied to the detectors. A $^{55}$Fe source hosted on the Nb magnetic shield is used to hit the detectors with 5.9~keV X-rays, typically with a count rate of $\sim1$~count per second per pixel.

The setup is housed in a Leiden Cryogenics dilution unit with a cooling power of 400~$\upmu$W at 120~mK. The setups are hung via Kevlar wires to the mixing chamber to damp mechanical oscillations\cite{gotkevlar}. OFHC copper braids connecting the setups to the mixing chamber ensure the thermal anchoring. The setup temperature is controlled via a Ge thermistor anchored to the copper holder and kept stable to 55~mK.

To characterize the gain sensitivities, we bias the pixels at a reference setpoint along the superconducting transition, defined by the reference values $V_0, T_0, B_0$ (where $B_0$ is chosen to minimize the residual magnetic field and $V_0$ corresponds to $R\approx 0.1 R_N$, where the best single pixel spectral performances are observed, likely due to the higher loop gain than at larger $R/R_N$), and at different setpoints obtained by individually changing each parameter by a quantity $\Delta V, \Delta T, \Delta B$, respectively. For each setpoint we acquire X-rays events with $\approx$ 500 events per pixel, in multiplexing mode. The X-ray energy of each event is assessed by using the X-ray pulse and noise information with the optimal filtering technique. To do so, an optimal filter template is generated for each pixel at the reference setpoint. To extract the impact on the estimated X-ray energy, each dataset is analysed using the optimal filter template of the reference setpoint. The energy scale is calibrated using the known K$\upalpha_1$ line of the $^{55}$Fe source spectrum, with energy $E_0 = 5898.75$~eV. To assess the energy for the sensitivity estimation, the acquired photons in the K$\alpha$ energy range are collected into a histogram and a gaussian fit is performed to extract the position $E$ for the K$\upalpha_1$ line. In this way, for each setpoint we can measure the shift in energy $\Delta E = E - E_0$ of the K$\upalpha_1$ line caused by the variation of the TES gain.

Repeating this action for several setpoints, we can fit $\Delta E$ as a function of $\Delta V, \Delta T, \Delta B$, respectively, to deduce a dependency and estimate the gain sensitivity for our TES arrays.

\section{Results and discussion}

In Table~\ref{tabres} we summarize the results of our gain sensitivity measurements, with a comparison with the requirements for Athena X-IFU. Since for X-IFU the requirements are at an energy of 7~keV, we also linearly scale up our values, measured with 5.9~keV photons.

From the IV curves we calibrate the reference voltage $V_0$ to bias each pixel at a resistance $R\approx 0.1 R_N$. We performed measurements also at higher bias points, up to $\approx 0.3 R_N$. Sensitivity values measured at such bias points are still compatible with what described in the following.

\begin{figure*}[!h]
\begin{center}
\subfigure[]{\includegraphics[width=0.49\textwidth]{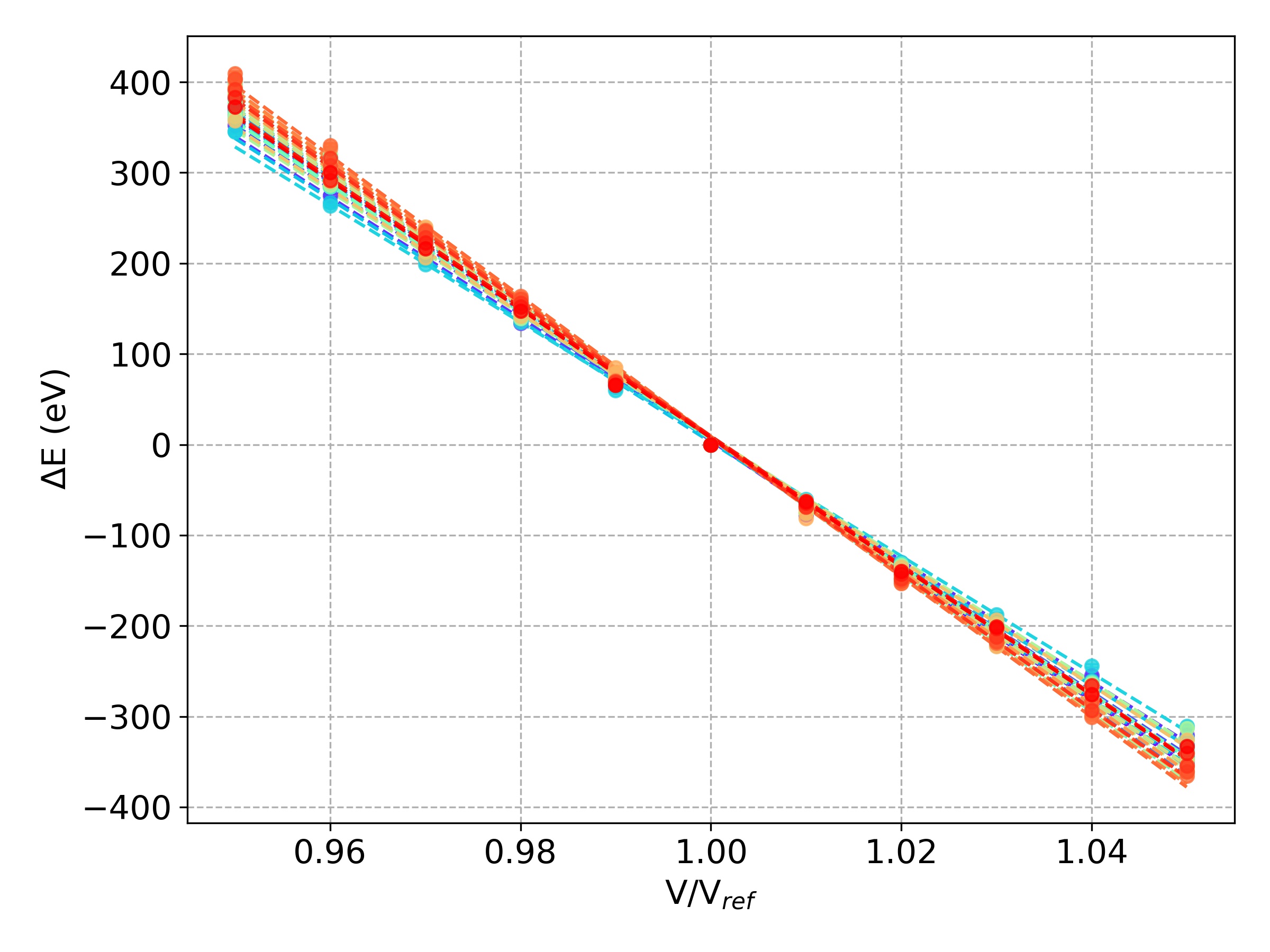}}
\subfigure[]{\includegraphics[width=0.49\textwidth]{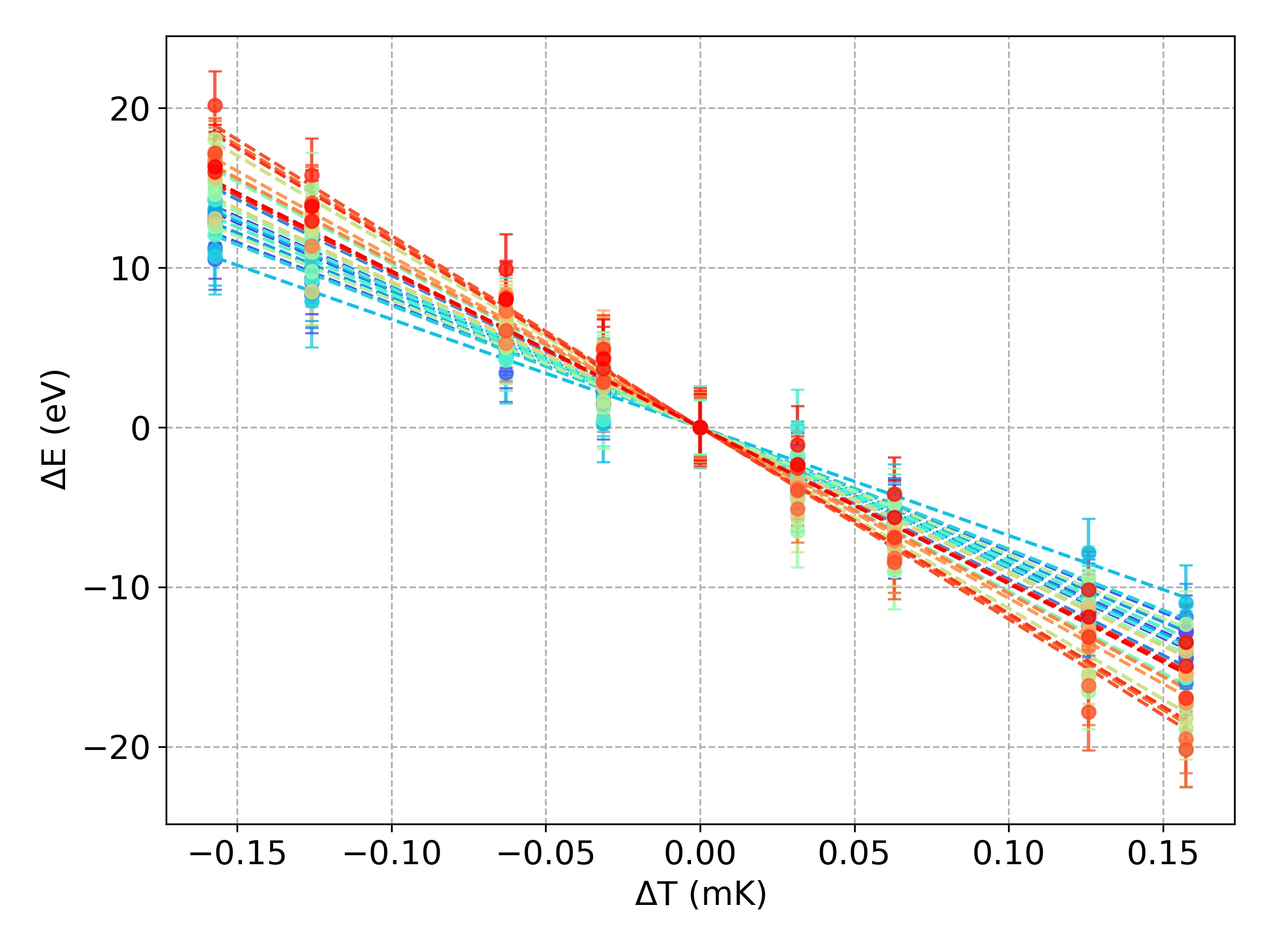}}\\
\subfigure[]{\includegraphics[width=0.49\textwidth]{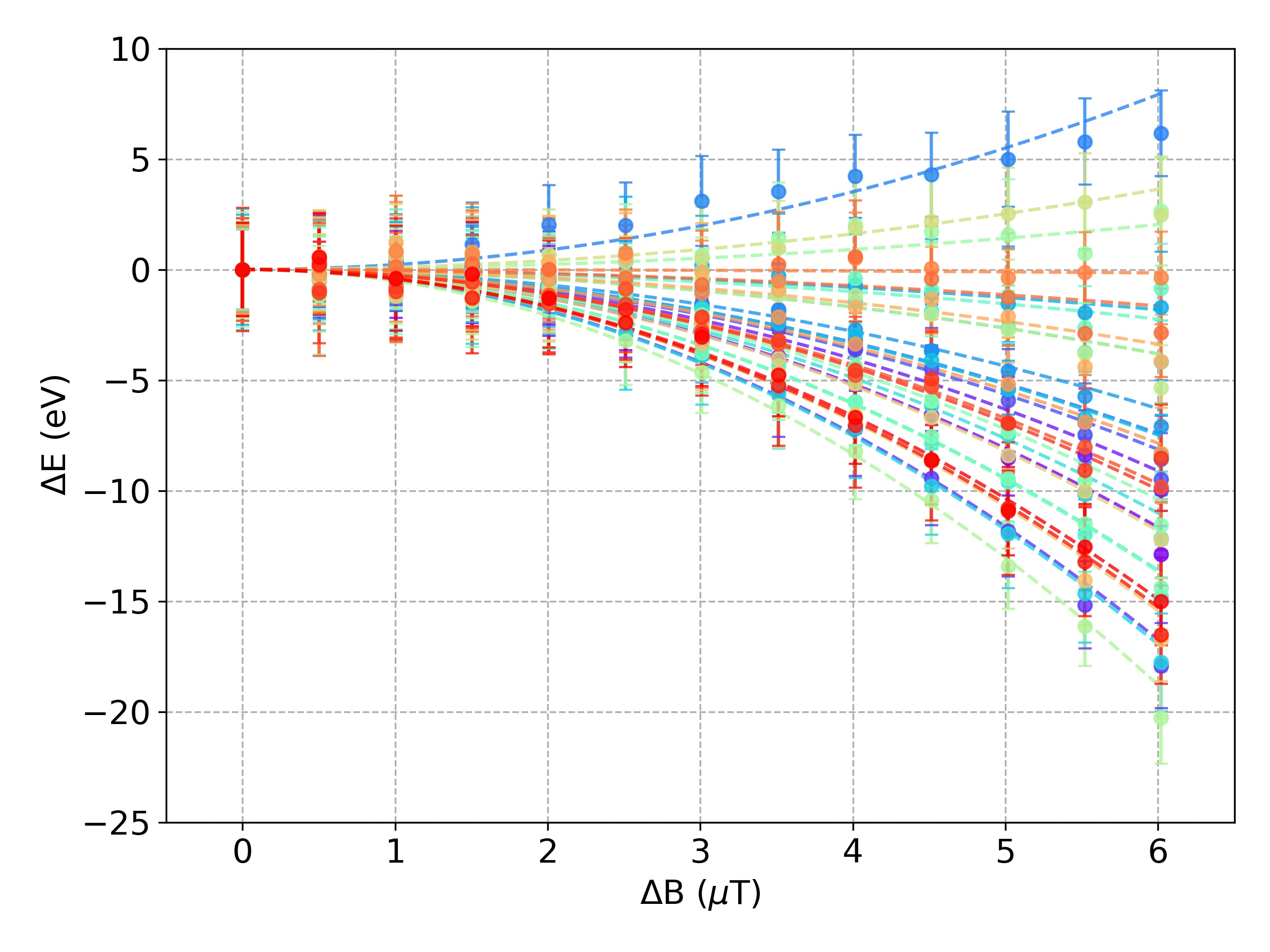}}
\subfigure[]{\includegraphics[width=0.49\textwidth]{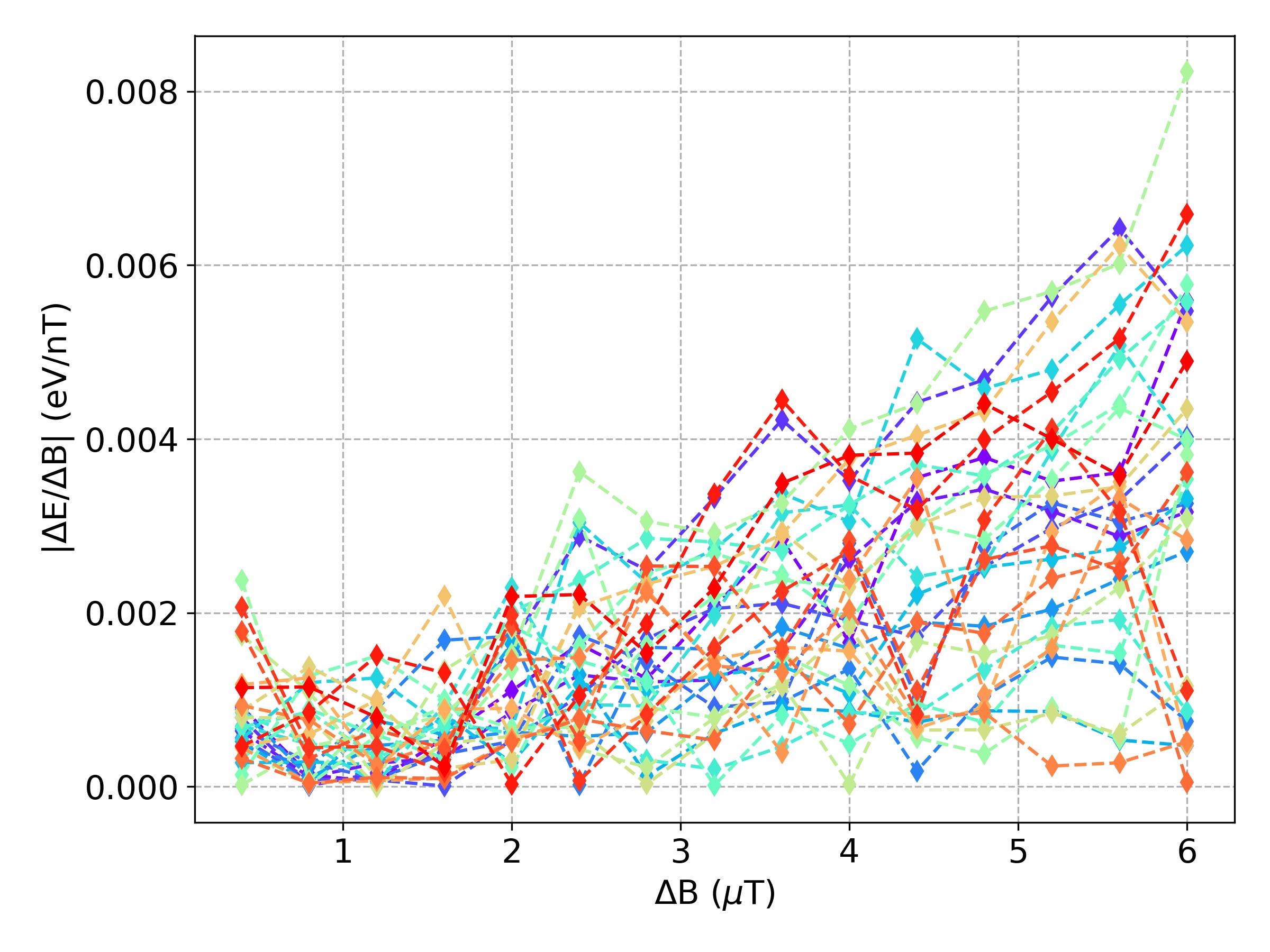}}\\
\caption{Measured sensitivity curves (each color represents a different pixel): (a) bias voltage sensitivity, (b) bath temperature sensitivity, (c) magnetic field sensitivity with (d) derivative of a linear interpolation of the data to extract the sensitivity from the non-linear dependency. (Color figure online).}\label{results}
\end{center}
\end{figure*}

To characterize the voltage sensitivity, we change the bias voltage of each pixel in a range $\Delta V = \pm 5\% = \pm 5\cdot10^4$~ppm. For each pixel we plot the shift in energy $\Delta E$ as a function of $\Delta V$ and use a linear fit to extract the dependency. The measured voltage sensitivity curves for each pixel are reported in Fig.~\ref{results}a. The results of the fit range from $6.4 \pm 0.8$~meV/ppm to $7.7 \pm 1.0$~meV/ppm, with a mean sensitivity of $7.2 \pm 0.3$~meV/ppm. 

We repeat the same process for the temperature sensitivity, varying the bath temperature in a range $\Delta T = \pm 160\ \upmu$K from a base temperature $T_0 = 55$~mK. The measured temperature sensitivities curves for each pixel are reported in Fig.~\ref{results}b. The results of the fit range from $67 \pm 3$~meV/$\upmu$K to $121 \pm 4$~meV/$\upmu$K, with a mean sensitivity of $94 \pm 14$~meV/$\upmu$K. The T-sensitivity values show a larger spread than the V-sensitivity case: we interpret this as $\alpha$ and $\beta$ not being exactly the same for pixels biased at different frequencies, as a consequence of the weak-link effect. For this geometry, at $R \approx 0.1R_N$ typical values for $\alpha$ and $\beta$ are 500 and 5, respectively.

To characterize the magnetic field susceptibility, we change the external magnetic field in a range $\Delta B = 6\ \upmu$T, after calibrating the starting value $B_0$ as the external magnetic field that in average minimizes the residual magnetic field for all the pixels. We only scan for positive $\Delta B$ values, since for our setup the sensitivity curve is symmetrical around the zero residual field $B_0$. In principle, this magnetic field dependence comes from the weak-link behaviour of the TES, acting as a SNS Josephson junction (S being the superconducting leads and N the TES bilayer itself)\cite{smithbfield} with a gauge-invariant phase $\varphi \propto \sqrt{PR}/\omega_0$ depending on the device power, resistance and the frequency of the magnetic field, either external or self-induced.

Fig.~\ref{results}c shows the measured magnetic field sensitivity curves, along with a fit using a second order polynomial. The different parabolic trends observed could be interpreted as a consequence of the frequency dependency of the TES weak-link behaviour. Given the non-linear behaviour, an univocal value for $\Delta E / \Delta B$ cannot be extracted as for the voltage and temperature sensitivities. Therefore, we perform a quadratic fit and calculate the sensitivity as the derivative of the fit function at a certain $\Delta B$.

In this way the sensitivities we obtain from the fit for $\Delta B = 100$~nT, where the X-IFU requirement is defined, are of the order of 1 meV/nT or less, but with errors of comparable value. To be conservative, we then estimate an upper limit, performing a linear interpolation of the data-points and directly calculating the differentials $dB$ and $dE$ as a function of the applied magnetic field. We then calculate the sensitivity as the derivative $dE/dB$. As shown in Fig.~\ref{results}d, across the measured range the sensitivity values are less than 10 meV/nT for all the pixels. At $\Delta B = 1\ \upmu$T, $i.e.$ the expected drift during one cool-down cycle on the X-IFU Focal Plane Assembly (FPA), the magnetic field sensitivity is $\lesssim 2$~meV/nT.

\begin{table*}[!h]
\begin{center}
    \begin{tabular}{ cccc }
    \textbf{Sensitivity} & \textbf{X-IFU req. @ 7 keV} & \textbf{Measured @ 5.9 keV} & \textbf{Scaled up to 7 keV} \\ \hline\hline
    $\Delta E/\Delta V$ & 15 meV/ppm & 7.2 meV/ppm $\pm$ 0.3 meV/ppm & $\approx$ 9 meV/ppm \\
    $\Delta E/\Delta T$ & 0.15 eV/$\upmu$K & 0.09 eV/$\upmu$K $\pm$ 0.01 eV/$\upmu$K & $\approx$ 0.1 eV/$\upmu$K \\
    $\Delta E/\Delta B$ & 8 eV/nT @ $0.1\ \upmu$T & $\lesssim 2\cdot10^{-3}$~eV/nT @ $1\ \upmu$T & $\lesssim 2\cdot10^{-3}$~eV/nT @ $1\ \upmu$T \\
    \hline
    \end{tabular}
\end{center}
\caption{Summary of the measured TES gain sensitivities. The requirements for X-IFU are derived by simulations with the $xifusim$\cite{xifusim} software, using an old TDM-optimized design of pixels.}\label{tabres}
\end{table*}

The voltage and temperature sensitivities are compliant with the X-IFU requirements within reasonable margin, the measured values for magnetic field sensitivity are orders of magnitude lower. Recent tests under dc readout, performed at NASA Goddard with a different pixel design, showed a magnetic field sensitivity at a level of 200~meV/nT\cite{smith2021}. Note that we are here considering the influence of dc magnetic fields and dc gradients on the array, since the main worry are magnetic fields at low frequencies such as stray fields from the cryo-cooling system on-board the satellite (compressor, ADR, etc.). This large difference with the magnetic field sensitivity measured with our TES arrays and FDM system can be understood as due to two concurrent factors.

The first one is the readout. Under ac-readout, the TES current is continuously sweeping between positive and negative values, hence intrinsically less sensitive to dc-magnetic fields than when readout using a dc-bias.

The second one is the TES design. The geometry of our TES bilayers for FDM readout has significantly evolved over the last years, moving from the classical large square, low-$R_N$ designs (more suitable for dc readout) towards smaller geometries with high aspect ratios\cite{martinhar} (width $W$ much smaller than the length $L$), resulting in higher normal resistances $R_N$, a feature more desirable for AC readout. In fact, as shown in Gottardi \emph{et al.}\cite{lgjosephson}, this optimization allows to minimize the weak-link effect, detrimental for the TES spectral performance under FDM readout.

Overall, the sensitivity then depends on a combination of TES design plus readout scheme. In principle, though optimal for ac-readout, higher $R_N$ devices could be used under dc-readout to mitigate the $B$-susceptibility. To our knowledge, the X-ray performance and $B$-sensitivity of such higher $R_N$ under dc-readout however has not yet been reported.

\begin{figure*}[!h]
\begin{center}
\subfigure[]{\includegraphics[height=0.35\textwidth]{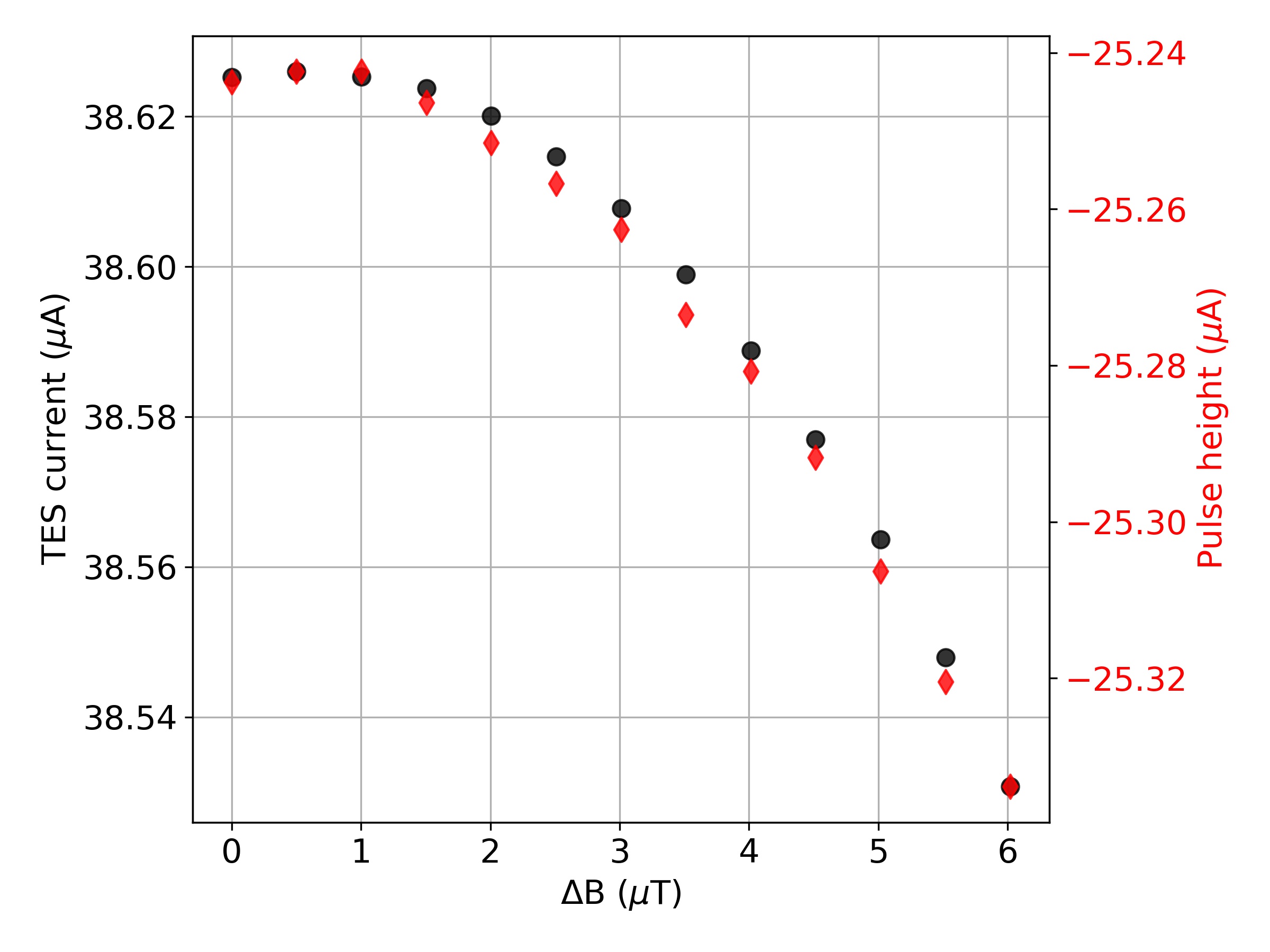}}
\subfigure[]{\includegraphics[height=0.37\textwidth]{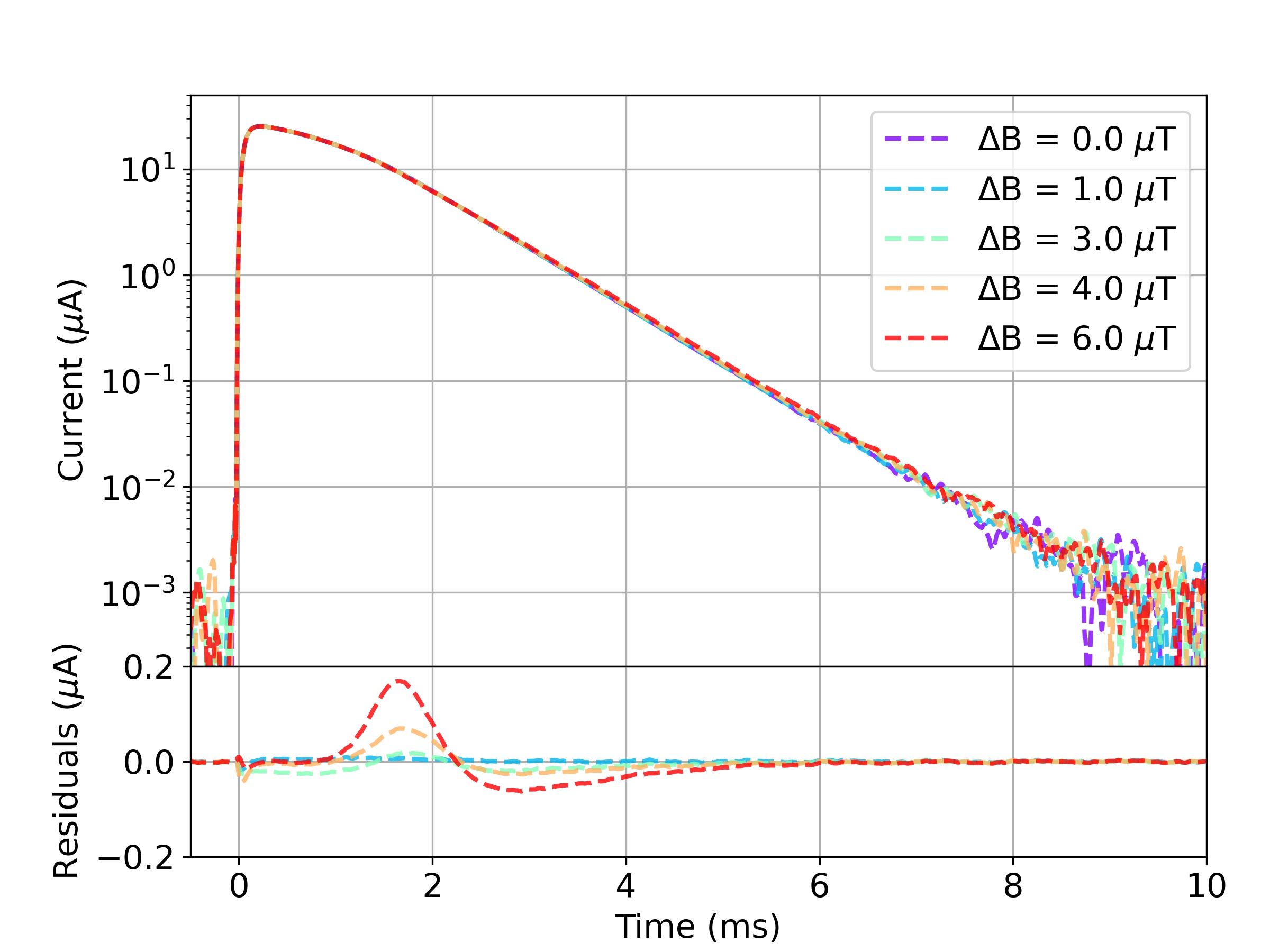}}\\
\caption{(a) Measured TES current (black) and pulse height (red) at different applied magnetic fields for the same pixel. The current is a good indicator for the pulse height, as no shift between the two curves is observed. (b) Comparison of X-ray pulses measured at different magnetic fields for the same pixel. The bottom panel refers to the residuals with the $B = 0\ \upmu$T data. (Color figure online).}\label{PHvsB}
\end{center}
\end{figure*}

In Fig.~\ref{PHvsB}a we show the measured TES current and the pulse height for 6~keV photons as a function of external magnetic field. As can be seen, there is no shift between the two curves which follow the same dependence on $B$, indicating that also the impact of the TES self-induced magnetic field is negligible\cite{smithbfield}. This means that the TES current can be used as a figure of merit for choosing the optimal setpoint to both minimize the $B$-sensitivity and maximize the pulse height, $i.e.$ without sacrificing energy resolution. Fig.~\ref{PHvsB}b shows that also the pulse shape is very well conserved at the different $\Delta B$, at a level better than 1\%.  

\begin{figure*}[!h]
\begin{center}
\subfigure{\includegraphics[trim={0 0 0 8mm},clip,width=0.32\textwidth]{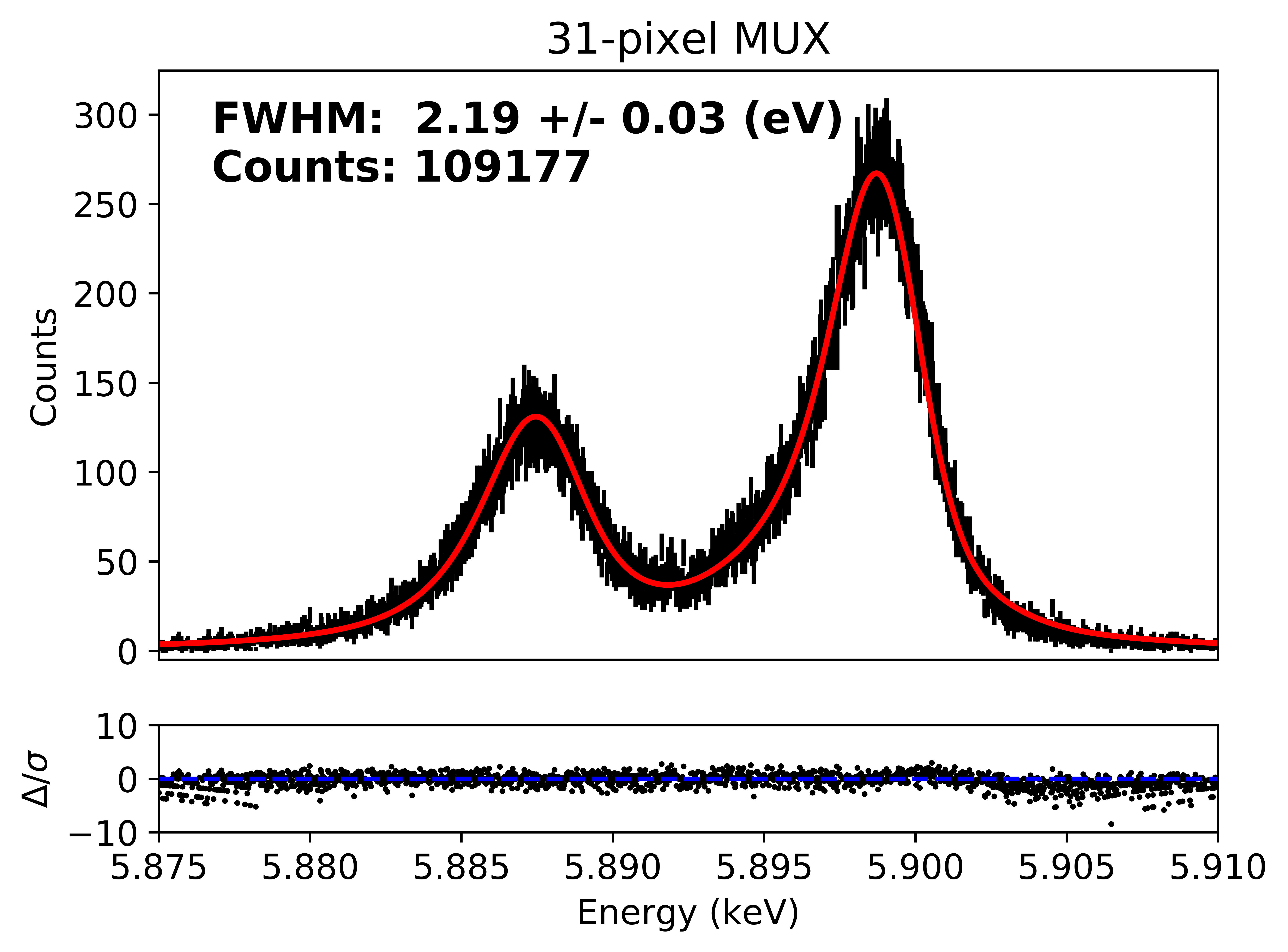}}
\subfigure{\includegraphics[trim={0 0 0 8mm},clip,width=0.32\textwidth]{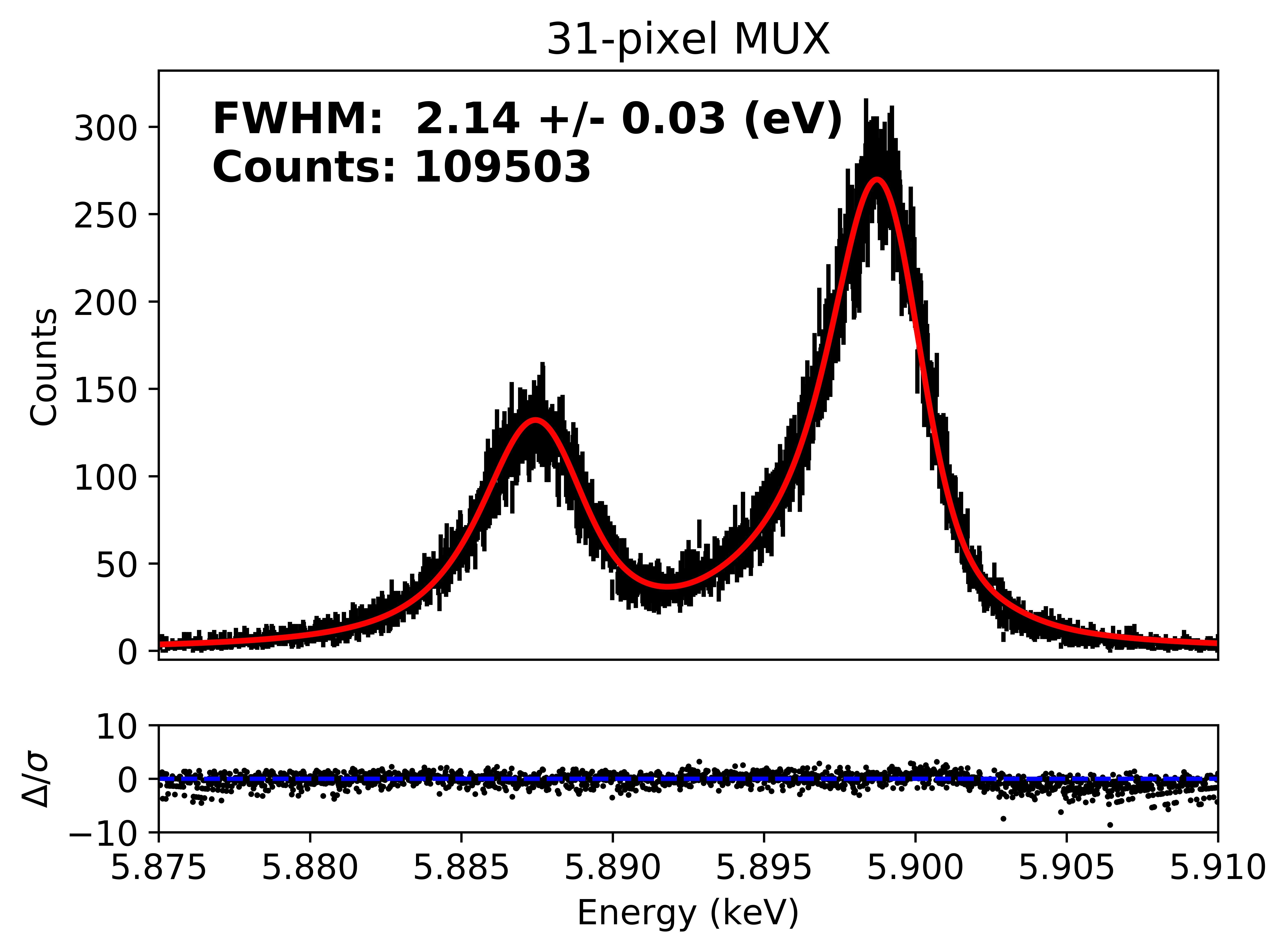}}
\subfigure{\includegraphics[trim={0 0 0 8mm},clip,width=0.32\textwidth]{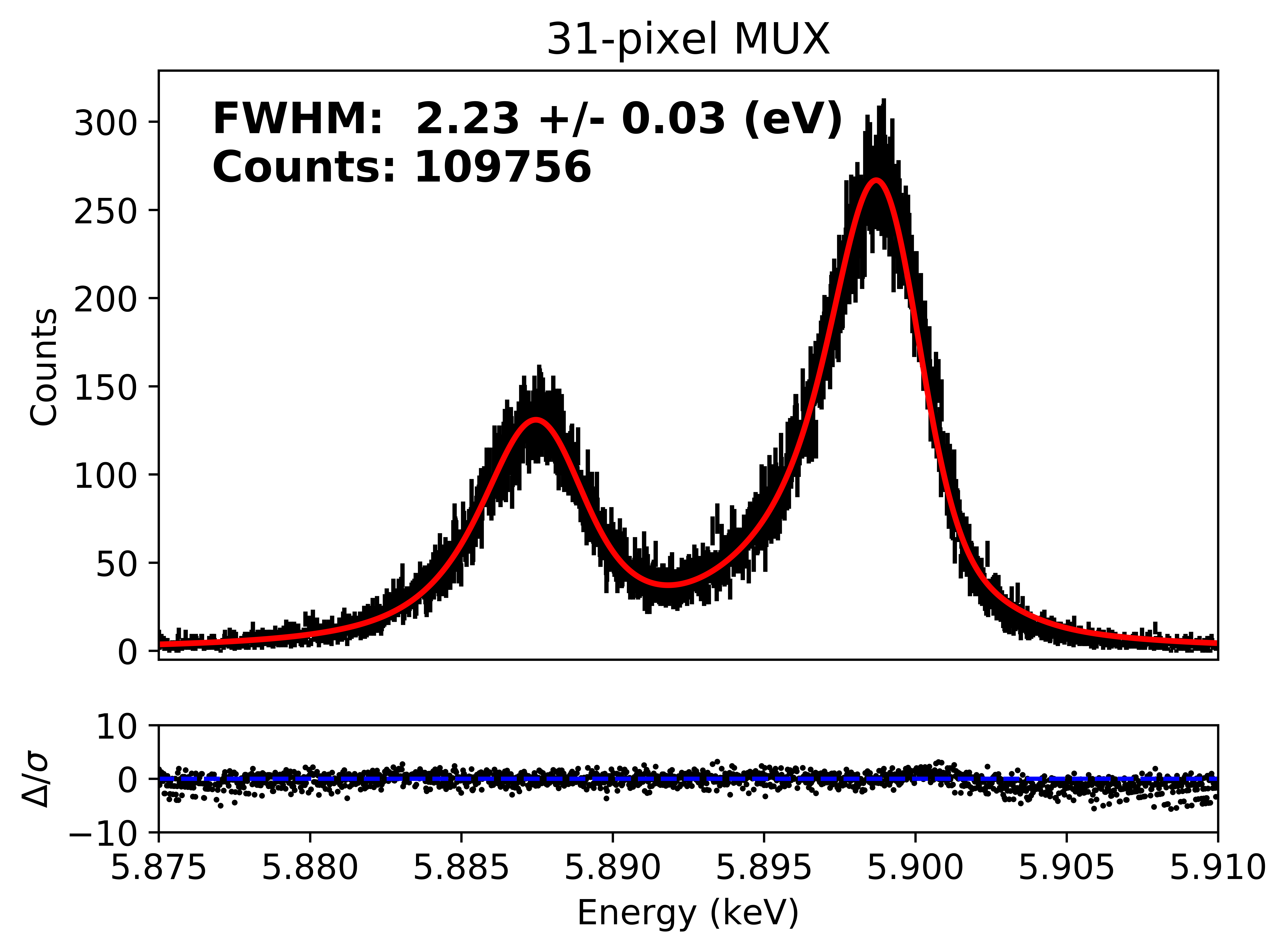}}\\
\caption{Energy resolutions measured with 5.9 keV photons at different values of applied magnetic field: $\Delta B = 0\ \upmu$T (left), $= -1\ \upmu$T (center), $= +1\ \upmu$T (right). (Color figure online).}\label{bres}
\end{center}
\end{figure*}

The measured $B$-field sensitivity should produce negligible impact on the energy resolution of the detectors for small changes in magnetic field. To verify this, we performed three consecutive X-ray measurements with all the 31 pixels active and simultaneously readout, for $\Delta B = 0$~$\upmu$T and $\pm 1$ $\upmu$T. Such values were chosen because 1 $\upmu$T is approximately the expected magnetic field gradient on the focal plane for X-IFU. The measured spectra, reported in Fig.~\ref{bres}, show consistent energy resolutions.

Such a negligible sensitivity to external dc magnetic fields of TES devices under FDM readout has important implications at an instrumental level: in particular, the design of the FPA of future space-borne instruments employing TES arrays could be greatly simplified if FDM readout is used, $e.g.$ with the reduction of the mass of the magnetic shielding.

\section{Summary}

We presented a characterization of TES gain sensitivities under FDM readout using 6~keV photons. The measured values are compliant with large margin with the requirements for the Athena X-IFU instrument. In particular, we found that TES geometries optimized for ac-readout have a sensitivity to dc-magnetic fields orders of magnitude lower than the current baseline under dc-bias for X-IFU. For future TES-based space-born missions, this would allow a simpler, lighter FPA design with much less stringent needs for magnetic shielding.

In the near future we plan to perform further measurements using a Modulated X-ray Source to probe different energies than 6~keV and with larger, kilo-pixel arrays, more representative of the actual arrays that would be used on a real instrument, to further validate these results and to verify the magnetic susceptibility of the energy scale calibration.

\section*{Acknowledgements}
SRON is financially supported by the Nederlandse Organisatie voor Wetenschappelijk Onderzoek.
This work is part of the research programme Athena with project number 184.034.002, which is (partially) financed by the Dutch Research Council (NWO).
The SRON TES arrays used for the measurements reported in this paper is developed in the framework of the ESA/CTP grant ITT AO/1-7947/14/NL/BW.

\section*{Data availability}

The corresponding author makes available the data presented in this paper upon reasonable request.


\begin{thebibliography}{99}

\bibitem{tes} L. Gottardi $and$ K. Nagayoshi, "A Review of X-ray Microcalorimeters Based on Superconducting Transition Edge Sensors for Astrophysics and Particle Physics", Applied Sciences 11(9), 3793 (2021).

\bibitem{athena} D. Barret, T. L. Trong, J.-W. den Herder, L. Piro, X. Barcons, J. Huovelin, R. Kelley, J. M. Mas-Hesse, K. Mitsuda, S. Paltani, \emph{et al.}, "The Athena X-ray Integral Field Unit (X-IFU)", Proc. SPIE 9905, Space Telescopes and Instrumentation 2016: Ultraviolet to Gamma Ray, 99052F (2016).

\bibitem{lynx} J. A. Gaskin, D. Swartz, A. A. Vikhlinin, F. Özel, K. E. E. Gelmis, J. W. Arenberg, S. R. Bandler, M. W. Bautz, M. M. Civitani, A. Dominguez \emph{et al.}, "Lynx X-Ray Observatory: an overview" , J. of Astronomical Telescopes, Instruments, and Systems, 5(2), 021001 (2019), DOI: 10.1117/1.JATIS.5.2.021001

\bibitem{hubs} W. Cui, L.-B. Chen, B. Gao, F.-L. Guo, H. Jin, G.-L. Wang, L. Wang, J.-J. Wang, W. Wang, Z.-S. Wang \textit{et al.}, "HUBS: Hot Universe Baryon Surveyor", Journal of Low Temperature Physics 199, 502–509 (2020).

\bibitem{tdm} M. Durkin \emph{et} al., "Demonstration of Athena X-IFU Compatible 40-Row Time-Division-Multiplexed Readout," IEEE Transactions on Applied Superconductivity, vol. 29, no. 5, pp. 1-5 (2019), DOI: 10.1109/TASC.2019.2904472.

\bibitem{hiroki2021} H.~Akamatsu, D.~Vaccaro, L.~Gottardi, J.~van~der~Kuur,~M.~Kiviranta, C.P.~de~Vries, M.P.~Bruijn, R.H.~den~Hartog, K.~Nagayoshi,  A.J.~van Linden \textit{et al.}, "Demonstration of 37-pixel frequency domain multiplexing readout of cryogenic X-ray microcalorimeters", Appl. Phys. Lett. 119, 182601 (2021).

\bibitem{ken} K. Nagayoshi, M. L. Ridder, M. P. Bruijn, L. Gottardi, E. Taralli, P. Khosropanah, H. Akamatsu, S. Visser $and$ J.-R. Gao, "Development of a Ti/Au TES Microcalorimeter Array as a Backup Sensor for the Athena/X-IFU Instrument", Journal of Low Temperature Physics 199, 943–948 (2020).

\bibitem{bbfb} Roland den Hartog, D. Boersma, M. Bruijn, B. Dirks, L. Gottardi, H. Hoevers, R. Hou, M. Kiviranta, P. de Korte, J. van der Kuur, \emph{et al.}, "Baseband feedback for frequency domain multiplexed readout of TES X-ray detectors", API Conference Proceedings 1185, 261 (2009).

\bibitem{marcel} M. P. Bruijn, L. Gottardi, R. H. den Hartog, J. van der Kuur, A. J. van der Linden $and$ B. D. Jackson, "Tailoring the high-Q LC filter arrays for readout of kilo-pixel TES arrays in the SPICA-Safari instrument", Journal of Low Temperature Physics, 176 421–425 (2014).

\bibitem{gotkevlar} L.~Gottardi, H.~van~Weers, J.~Dercksen, H.~Akamatsu,  M.~P.~Bruijn, J.~R.~Gao, B.~Jackson, P.~Khosropanah, J.~van~der~Kuur, K.~Ravensberg and M.~L.~Ridder, "A six-degree-of-freedom micro-vibration acoustic isolator for low-temperature radiation detectors based on superconducting transition-edge sensors", Review of Scientific Instruments 90, 055107 (2019).

\bibitem{smithbfield} Stephen J. Smith, Joseph S. Adams, Catherine N. Bailey, Simon R. Bandler, Sarah E. Busch, James A. Chervenak, Megan E. Eckart, Fred M. Finkbeiner, Caroline A. Kilbourne, Richard L. Kelley, Sang-Jun Lee, Jan-Patrick Porst, Frederick S. Porter, $and$ John E. Sadleir , "Implications of weak-link behavior on the performance of Mo/Au bilayer transition-edge sensors", Journal of Applied Physics 114, 074513 (2013) https://doi.org/10.1063/1.4818917

\bibitem{xifusim} M. Lorenz, C. Kirsch, P.E. Merino-Alonso \emph{et al.}, "GPU Supported Simulation of Transition-Edge Sensor Arrays", J Low Temp Phys 200, 277–285 (2020). https://doi.org/10.1007/s10909-020-02450-1

\bibitem{martinhar} M. de Wit, L. Gottardi, E. Taralli, K. Nagayoshi, M. L. Ridder, H. Akamatsu, M. P. Bruijn, M. D’Andrea, J. van der Kuur, K. Ravensberg, D. Vaccaro, S. Visser, J. R. Gao, and J.-W. A. den Herder , "High aspect ratio transition edge sensors for x-ray spectrometry", Journal of Applied Physics 128, 224501 (2020) https://doi.org/10.1063/5.0029669

\bibitem{lgjosephson} L. Gottardi, S. J. Smith, A. Kozorezov, H. Akamatsu, J. van der Kuur, S. R. Bandler, M. P. Bruijn, J. A. Chervenak, J. R. Gao, R. H. den Hartog, B. D. Jackson, P. Khosropanah, A. Miniussi, K. Nagayoshi, M. Ridder, J. Sadleir, K. Sakai and N. Wakeham, "Josephson Effects in Frequency-Domain Multiplexed TES Microcalorimeters and Bolometers", Journal of Low Temperature Physics volume 193, pages 209–216 (2018).

\bibitem{smith2021} S. Smith \emph{et al.}, "Influence of environmental parameters on the performance of transition edge sensors", Journal of Low Temperature Physics, This Special Issue (2021).

\end{thebibliography}
\end{document}